\documentclass[twocolumn,
superscriptaddress,
 amsmath,amssymb,
 aps,
 prl,
floatfix,
]{revtex4-2}

\usepackage{physics}
\usepackage{graphicx}
\usepackage{dcolumn}
\usepackage{bm}

\usepackage[breaklinks=true,colorlinks,citecolor=blue,linkcolor=blue,urlcolor=blue]{hyperref}

\newcommand{\sz}{\hat \sigma_z}

\newcommand{\aop}{\hat a}
\newcommand{\adop}{\hat a ^\dagger}

\newcommand{\figpanel}[2]{Fig.~\hyperref[#1]{\ref*{#1}(#2)}}
\newcommand{\figurepanel}[2]{Figure~\hyperref[#1]{\ref*{#1}(#2)}}
\newcommand{\figpanels}[3]{Figs.~\hyperref[#1]{\ref*{#1}(#2)-(#3)}}

\newcommand{\be}{\begin{equation}}
\newcommand{\ee}{\end{equation}}
\newcommand{\bea}{\begin{eqnarray}}
\newcommand{\eea}{\end{eqnarray}}

\renewcommand{\eqref}[1]{\mbox{Eq.~(\ref{#1})}}
\newcommand{\eqaref}[1]{\mbox{Equation~(\ref{#1})}}

\begin{document}

\title{Pure Dephasing of Light-Matter Systems \\in the Ultrastrong and Deep-Strong Coupling Regimes}

\author{Alberto Mercurio}
\affiliation{Dipartimento di Scienze Matematiche e Informatiche, Scienze Fisiche e  Scienze della Terra, Universit\`{a} di Messina, I-98166 Messina, Italy}

\author{Shilan Abo}
\email{shsavan@gmail.com}
\affiliation{Dipartimento di Scienze Matematiche e Informatiche, Scienze Fisiche e  Scienze della Terra, Universit\`{a} di Messina, I-98166 Messina, Italy}
\affiliation{Institute of Spintronics and Quantum Information, Adam Mickiewicz University, 61-614, Poznan, Poland}

\author{Fabio Mauceri}
\email{fabio.mauceri@unime.it}
\affiliation{Dipartimento di Scienze Matematiche e Informatiche, Scienze Fisiche e  Scienze della Terra, Universit\`{a} di Messina, I-98166 Messina, Italy}


\author{Enrico Russo}
\affiliation{Dipartimento di Scienze Matematiche e Informatiche, Scienze Fisiche e  Scienze della Terra, Universit\`{a} di Messina, I-98166 Messina, Italy}

\author{Vincenzo Macr\`{i}}
\affiliation{Theoretical Quantum Physics Laboratory, RIKEN Cluster for Pioneering Research, Wako-shi, Saitama 351-0198, Japan}

\author{Adam Miranowicz}
\affiliation{Institute of Spintronics and Quantum Information, Adam Mickiewicz University, 61-614, Poznan, Poland}

\author{Salvatore Savasta}

\affiliation{Dipartimento di Scienze Matematiche e Informatiche, Scienze Fisiche e  Scienze della Terra,
	Universit\`{a} di Messina, I-98166 Messina, Italy}

\author{Omar Di Stefano}
\affiliation{Dipartimento di Scienze Matematiche e Informatiche, Scienze Fisiche e  Scienze della Terra, Universit\`{a} di Messina, I-98166 Messina, Italy}

\date{\today}

\begin{abstract}
Pure dephasing originates from the non-dissipative information exchange between quantum systems and environments, and plays a key-role in both  spectroscopy and quantum information technology. Often pure dephasing constitutes the main mechanism of decay of quantum correlations. Here we investigate how pure dephasing of one of the components of a hybrid quantum system affects the dephasing rate of the system transitions. We find that, in turn, the interaction, in the case  of a light-matter system, can significantly affect the form of the stochastic perturbation  describing the dephasing of a subsystem, depending on the adopted gauge. Neglecting this issue can lead to wrong and unphysical results when the interaction becomes comparable to the bare resonance frequencies of subsystems, which correspond to the ultrastrong and deep-strong coupling regimes. We present results for two prototypical models of cavity quantun electrodynamics: the quantum Rabi and the Hopfield model.
\end{abstract}

\maketitle
{\em Introduction.---}In reality, there is no perfectly isolated quantum system. For example, the coupling of a radiating atom with the infinitely many modes of a free electromagnetic field results in decoherence and spontaneous emission. Such interaction determines an energy relaxation time $T_1$ associated to a given optical transition. If the population of an excited state decays, so does the polarization too, which results in decoherence. In the presence of only energy relaxation mechanisms, such transverse relaxation time is $T_2  = 2 T_1$ \cite{Allen1987,lambropoulos2007fundamentals}. However, quantum systems, displaying optical transitions, do not only interact with an electromagnetic field, but can be affected by additional dephasing mechanisms inducing the decay of the dipole coherence without changing the populations of the systems. 
These pure dephasing effects can originate from fluctuations in the environmental fields affecting the phases of the emitter wave functions. For example, dipoles in a medium also interact with phonons of the host lattice, they might collide with each other in a gas laser, and so on \cite{Borri2001UltralongDephasing, borri2003exciton, Muljarov2005Phonon, Delbecq2016QuantumDephasing, Kilen2020Propagation, Katsch2020Exciton}. Pure dephasing effects can be described by virtual
processes, which start from a relevant state and, after some
excursion in the intermediate states, return to the same initial
state.
In general, the phase (transverse) relaxation time is most often shorter than twice the energy relaxation time: $T_2 \leq 2 T_1$. In optical spectroscopy, the full width at half maximum (FWHM) of homogeneous broadening  corresponds to  $2/T_2$.

It is well known that decoherence tends to destroy quantum coherence and quantum correlations \cite{Eberly2003, Haroche2013}. It is known that this mechanism becomes faster with the increase of the {\em size} of a quantum system \cite{Zurek2003Decoherence}. This explains the absence of quantum superpositions in the macroscopic world \cite{Habib1998}.  
Decoherence can, thus, strongly affect and limit quantum information processing (QIP) \cite{DiVincenzo1995QuantumComputation, Simeon2019Overcoming}. Depending on the specific environment, mechanisms to protect qubits from dephasing have been proposed (see, e.g., \cite{Lidar1998,Grodecka2006,Ban2006,Guo2018, Simeon2019Overcoming}).

\begin{figure}[b]
    \centering
    \includegraphics[width = 0.9 \linewidth]{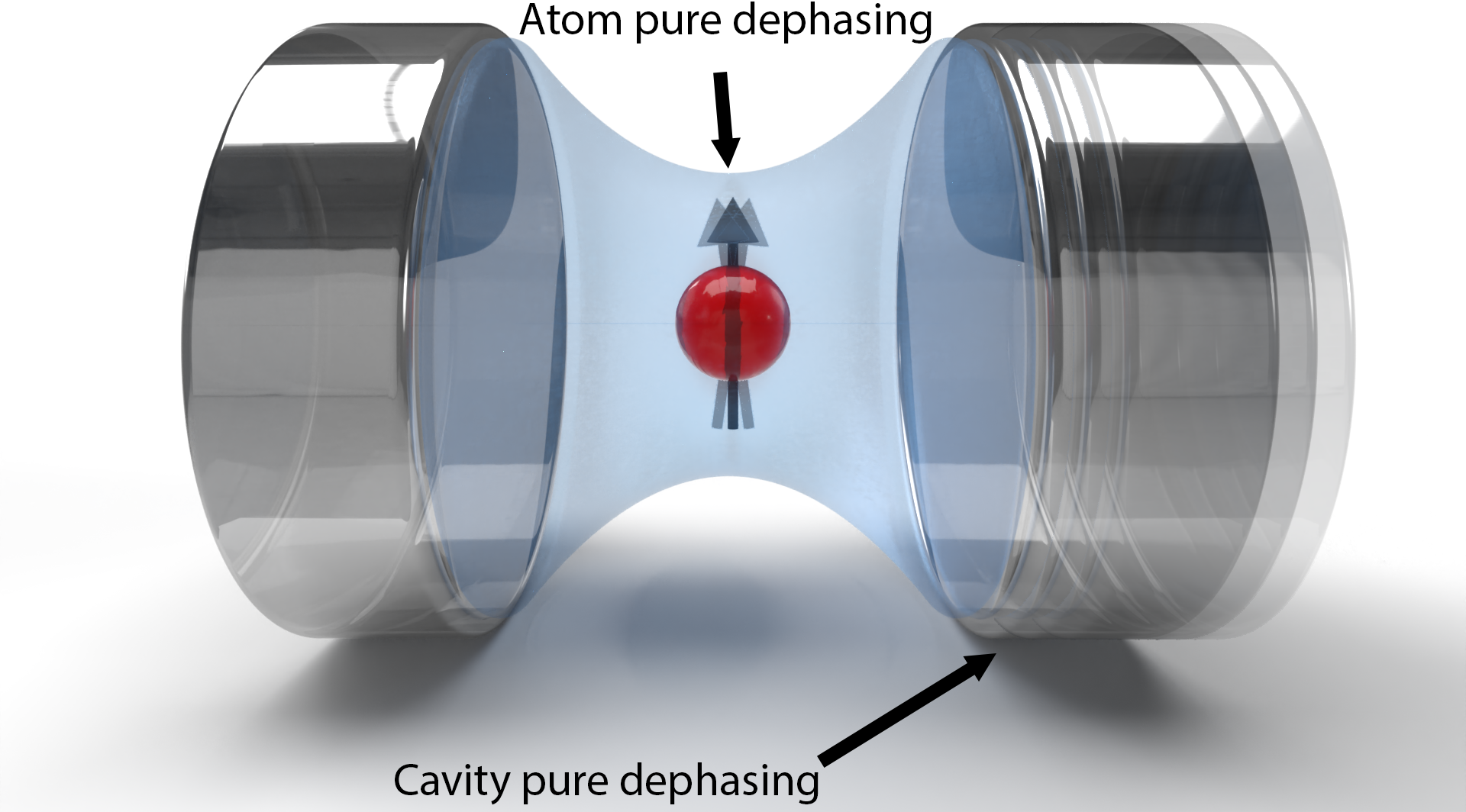}
    \caption{Pictorial representation of a two-level system interacting with a single-mode cavity field, when both subsystems are affected by pure dephasing.}
    \label{fig:pictorial_representation}
\end{figure}

Devices for QIP, secure communication, and high-precision sensing were implemented combining different systems ranging from photons, atoms, and spins to mesoscopic superconducting and nanomechanical structures. Complementary functionalities of these  hybrid quantum systems can be essential for the development of new quantum  technologies \cite{Xiang2013Hybrid, kurizki2015quantum, clerk2020hybrid}. Understanding how decoherence of one or more subsystems can affect the performance of the whole system is an interesting problem, relevant for improving the performance of quantum devices \cite{Beaudoin2011,Settineri2018}.

Cavity \cite{Haroche2013} and circuit \cite{Xiang2013, Gu2017} quantum electrodynamics (QED) systems are among the most studied hybrid quantum systems. 
They are playing a key role in quantum optics and in the development of new quantum technologies \cite{Lounis2005Single, gambetta2017building, Ganzhorn2019Gate, Uppu2020Scalable}. Pure dephasing can significantly affect the performance of these systems, not necessarily in a negative way.  
For example, it has been shown that pure dephasing is a promising resource for solid-state emitters, since it can improve the performance of nanophotonic devices, such as single-photon sources and nanolasers \cite{Auffeves2010}.

Decoherence effects in hybrid quantum systems are often introduced by using the standard quantum optics master equation, where the coupling of a multicomponent system with the environment is introduced by neglecting the interaction between the subsystems. When such interaction is not negligible compared to the bare transition frequencies of the components, as in the light-matter ultrastrong coupling (USC) or deep-strong coupling (DSC) regimes \cite{Kockum2018, Forn-Diaz2018}, this approximation can give rise to unphysical results. A master equation method, fully taking into account matter-light interaction, has been proposed in \cite{Beaudoin2011, Settineri2018}. These regimes can give rise to new physical effects and applications (see, e.g., \cite{Ma2015, Garziano2015,Kyaw2015, Garziano2016, Stassi2018, Stassi2020}), and they also challenge our understanding of fundamental aspects of cavity QED, like a proper definition of subsystems, their quantum measurements, the structure of the light-matter ground states, leading also to gauge ambiguities \cite{DeBernardis2018,DiStefano2019, Taylor2020, Dmytruk2020, Settineri2021, Stokes2021,Savasta2021}.

Here we show that the interaction between light and matter can significantly affect the form of a stochastic perturbation  describing the dephasing of one of the components, depending on the adopted gauge. We find that, neglecting this issue can lead to wrong and unphysical results in both USC and DSC regimes. We present results for two prototypical models of cavity QED: the quantum Rabi model (QRM) and the Hopfield model.

{\em Quantum Rabi model.---}Pure dephasing effects on a qubit can be described by introducing a zero-mean stochastic function $f(t)$ modulating its transition frequency. The perturbation Hamiltonian is 
\be\label{ClassicalDephasingNoise}
\hat {\cal V}^q_{\rm dep}= f_q(t) \sz\, .
\ee
When the qubit is a component of a hybrid quantum system, by expanding $\hat V^q_{\rm dep}$ in the basis of the eigenstates of the total system Hamiltonian, a master equation describing the effects of qubit dephasing on the system dynamics can be obtained \cite{Beaudoin2011}. For the sake of simplicity, we consider a stochastic function with a low-frequency spectral density (with respect to the relevant transition frequencies of the system). The resulting master equation can be written as  ($\hbar = 1$) \cite{Breuer2002}:
\be\label{Mastereq0}
\frac{d}{dt}\hat{\rho}(t)=-i[\hat H_s, \hat{\rho}]+ \frac{\gamma^0_\phi}{2}{\cal D}[\hat \Phi] \hat{\rho}\, ,
\ee
where $\hat H_s$ is the Hamiltonian of the total system and
\be
{\cal D}[\hat O] \hat{\rho} = \frac{1}{2} ( 2\hat O \hat{\rho} \hat O^\dag - \hat{\rho} \hat O^\dag \hat O - O^\dag \hat O \hat{\rho})\,
\ee
is the Lindbladian superoperator, while $\hat \Phi = \sum_j \hat \sigma_z^{jj} |j\rangle \langle j |$, with $|j \rangle$ being the eigenstates of $\hat H_s$, and $\sigma_z^{jj} = \langle j| \sigma_z | j \rangle$. The bare dephasing rate $\gamma^0_\phi = 2 S_f(0)$ is determined by the low-frequency spectral density $S_f(\omega)$ of $f(t)$.
Additional dephasing terms can appear, when the spectral density function $S_f(\omega)$ is not negligible at the transition frequencies of the system (see the Supplemental Material \cite{SupMat}).

We apply the above procedure to the simplest model of cavity-QED, i.e. the QRM. Its Hamiltonian in the dipole gauge can be written as $\hat{\mathcal{H}}_{D}= \hat{\mathcal{H}}_{\rm ph} + \hat{\mathcal{H}}_q + \hat{\mathcal{V}}_D$, where $\hat{\mathcal{H}}_q={\omega_q}\hat{\sigma}_z/2$, and the free field Hamiltonian is $\hat{\mathcal{H}}_{\rm ph}=\omega_c \hat{a}^{\dagger}\hat{a}$, where $\hat \sigma_j$ ($j = x,y,z$) are the Pauli operators, and $\hat a$ and $\hat a^\dag$ are the photon destruction and creation operators. Neglecting the constant term $\eta^2 \omega_c$, the interaction term can be written as $\hat{\mathcal V}_D = - i \eta \omega_c \left(\hat{a}-\hat{a}^{\dagger}\right) \hat{\sigma}_{x}$,
where $\eta$ is the normalized qubit-cavity coupling strength.
The correct Coulomb-gauge quantum Rabi Hamiltonian \cite{DiStefano2019} differs from the standard one, which violates gauge invariance, and it reads
\begin{equation}\label{eq:H Coulomb}
\hat{\mathcal{H}}_{C} =
\hat{\mathcal{H}}_{\rm ph}
+ \frac{ \omega_{q}}{2}
\left[\hat{\sigma}_{z} \cos (2 \hat {\cal A}) + \hat{\sigma}_{y} \sin (2 \hat {\cal A}) \right]\, ,
\end{equation}
where $\hat {\cal A} = \eta (\hat a + \hat a^\dag)$.
The dipole and Coulomb gauge Hamiltonians are related by the unitary gauge transformation: $\hat{\cal{H}}_{D} = \hat {\cal T} \hat{\cal{H}}_{C} \hat {\cal T}^\dag$, where $\hat {\cal T} = {\rm exp}[-i \hat {\cal A} \hat \sigma_x]$.

Pure dephasing effects can be directly introduced by using \eqref{Mastereq0}, which provides gauge invariant expectation values, as can be easily shown \cite{Mercurio2022Incoherent}. However, this is not sufficient to ensure that the obtained results are physically correct. In particular, we show below that, if one uses  $\hat {\cal H}_s = \hat {\cal H}_D$ or $\hat {\cal H}_s = \hat {\cal H}_C$  in \eqref{Mastereq0}, significantly different results can be obtained. 

To clarify this subtle but relevant point, we observe that different gauges can determine different expressions of the physical observables of the light and matter components \cite{Cohen-Tannoudji1997}. These differences become more relevant in the USC regime. For example, in the Coulomb-gauge the atom momentum is affected by light-matter interaction, while in the dipole gauge it is interaction independent. On the contrary, the dipole gauge affects the definition of the field momenta. Since, in the dipole gauge, all the atomic canonical variables are unaffected by interactions, we can safely describe pure dephasing effects starting from  the same potential as for the non-interacting case in \eqref{ClassicalDephasingNoise}. Notice that the population difference operator $\hat \sigma_z$, in \eqref{ClassicalDephasingNoise}, is not gauge invariant. As a consequence, using \eqref{ClassicalDephasingNoise}, to describe pure dephasing effects in the Coulomb gauge, gives different and wrong results.
However, once clarified this point, correct results can be obtained in any gauge. For example, the bare  $\hat \sigma_z$ operator in \eqref{ClassicalDephasingNoise}, when using the Coulomb gauge, becomes $\hat \sigma_z^C = \hat {\cal T}^\dag \hat \sigma_z \hat {\cal T}$. After that, correct gauge invariant results can also be obtained in the Coulomb gauge. 
At least in principle, pure dephasing effects can also affect the cavity field. In this case a perturbation potential of the form $\hat V^c_{\rm dep} = f_c(t) \hat a^\dag \hat a$ can only be used in the Coulomb gauge, where the field canonical variables are not affected by interactions.
In general, the form of the pure-dephasing perturbation potential is gauge dependent. In the presence of both dephasing channels, it can be written as
\be
\label{eq: V_dep in Coulomb}
\hat {\cal V}^C_{\phi} = f_q(t)  \hat \sigma^C_z + f_c(t) \hat a^\dag \hat a\, ,
\ee
in the Coulomb gauge, and as 
\be
\label{eq: V_dep in dipole}
\hat {\cal V}^D_{\phi} = f_q(t)  \hat \sigma_z + f_c(t) \hat a_D^\dag \hat a_D\, ,
\ee
in the dipole gauge, where $\hat a_D = \hat {\cal T} \hat a \hat {\cal T}^\dag = \hat a  + i \eta \hat \sigma_x$.
\begin{figure}[t]
    \centering
    \includegraphics[width= 1\linewidth]{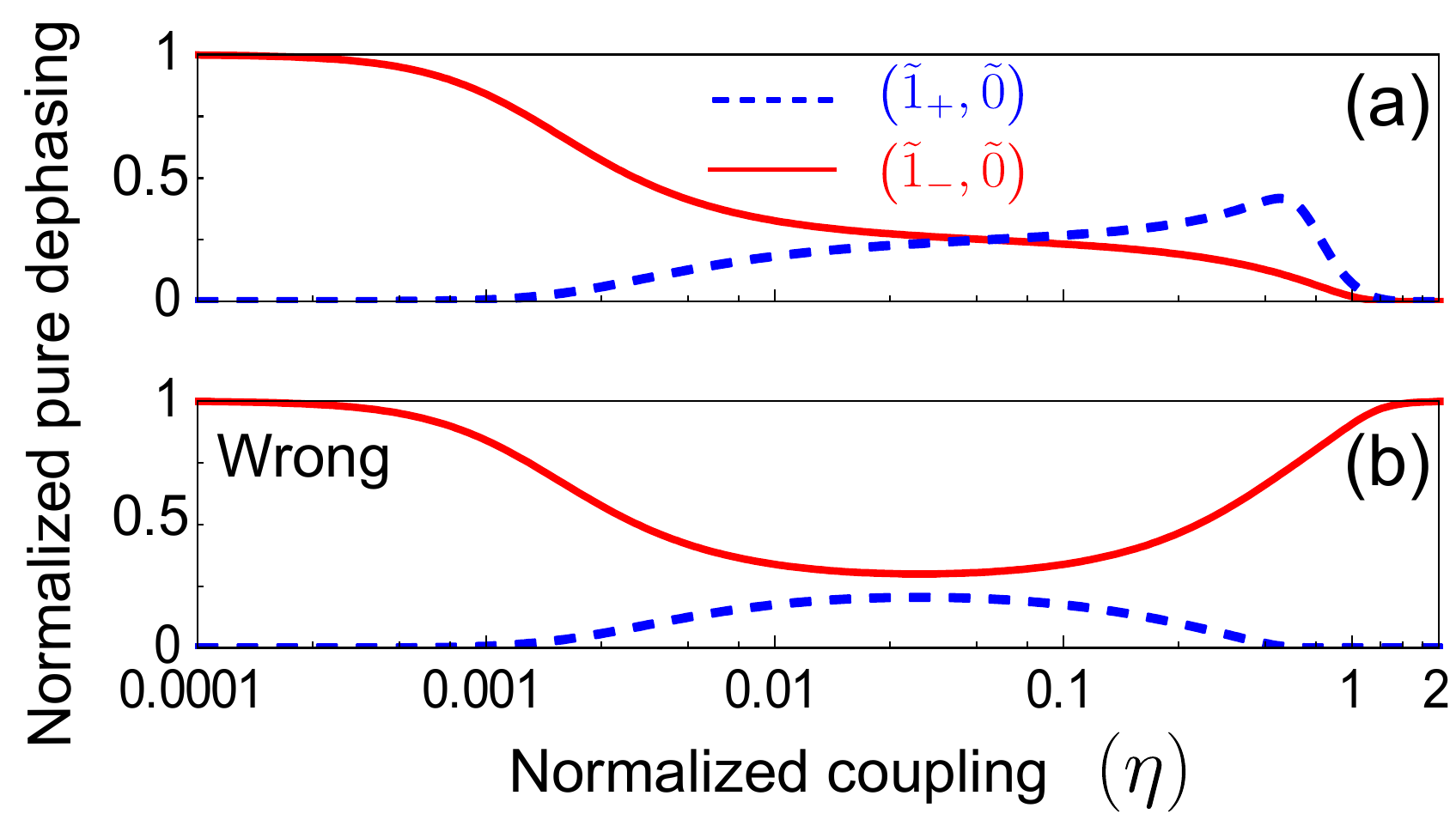}
    \caption {Quantum Rabi Model. Normalized pure dephasing rate 
    for the two lowest energy transitions, for a small qubit-cavity detuning $\delta = 3 \times 10^{-3}$ and considering only the qubit pure dephasing. (a) Correct gauge-invariant versus (b) wrong Coulomb gauge results.}
    \label{fig:Cav_stra}
\end{figure}

In the following, we label the QRM states by generalizing the notation of the Jaynes-Cummings (JC) model. In particular, $| \tilde 0 \rangle$ denotes the ground state, and $| \tilde n_\pm \rangle$ denote the states that tend to the JC states $| n_\pm \rangle$, when the coupling vanishes. Moreover, we use not-primed (primed) states to indicate the Coulomb (dipole) gauge states.

We now analyze pure dephasing effects on the two lowest transitions in the QRM: $\alpha_\pm \equiv (\tilde 1_\pm, \tilde 0)$.  In the interaction picture, from \eqref{Mastereq0}, we obtain \cite{SupMat}:
\be
\dot {\tilde \rho}_{\alpha'_\pm}(t) = - \left( \gamma_\phi^{\alpha'_\pm}/2 \right) \tilde \rho_{\alpha'_\pm}(t)\,,
\ee
with 
\be \label{puredqubit}
\gamma_\phi^{\alpha'_\pm} = \frac{\gamma^0_\phi}{2} \left| \sigma_z^{\tilde 1'_\pm,\tilde 1'_\pm} - \sigma_z^{\tilde 0',\tilde 0'}\right|^2\, .
\ee
Notice that here the expectation values $\sigma_z^{j',j'} \equiv \langle j'| \hat \sigma_z |j' \rangle$ have to be calculated using the dipole gauge. Of course, the obtained dephasing rate is gauge invariant ($\gamma_\phi^{\alpha'_\pm} = \gamma_\phi^{\alpha_\pm}$), because the expectation values are unitary invariant, when transforming both  operator and states: $\gamma_\phi^{\alpha_\pm} = {\gamma^0_\phi} | \sigma_z^{C,\tilde 1_\pm,\tilde 1_\pm} - \sigma_z^{C,\tilde 0,\tilde 0}|^2/2$.

\begin{figure}[b]
    \centering
    \includegraphics[width=\linewidth]{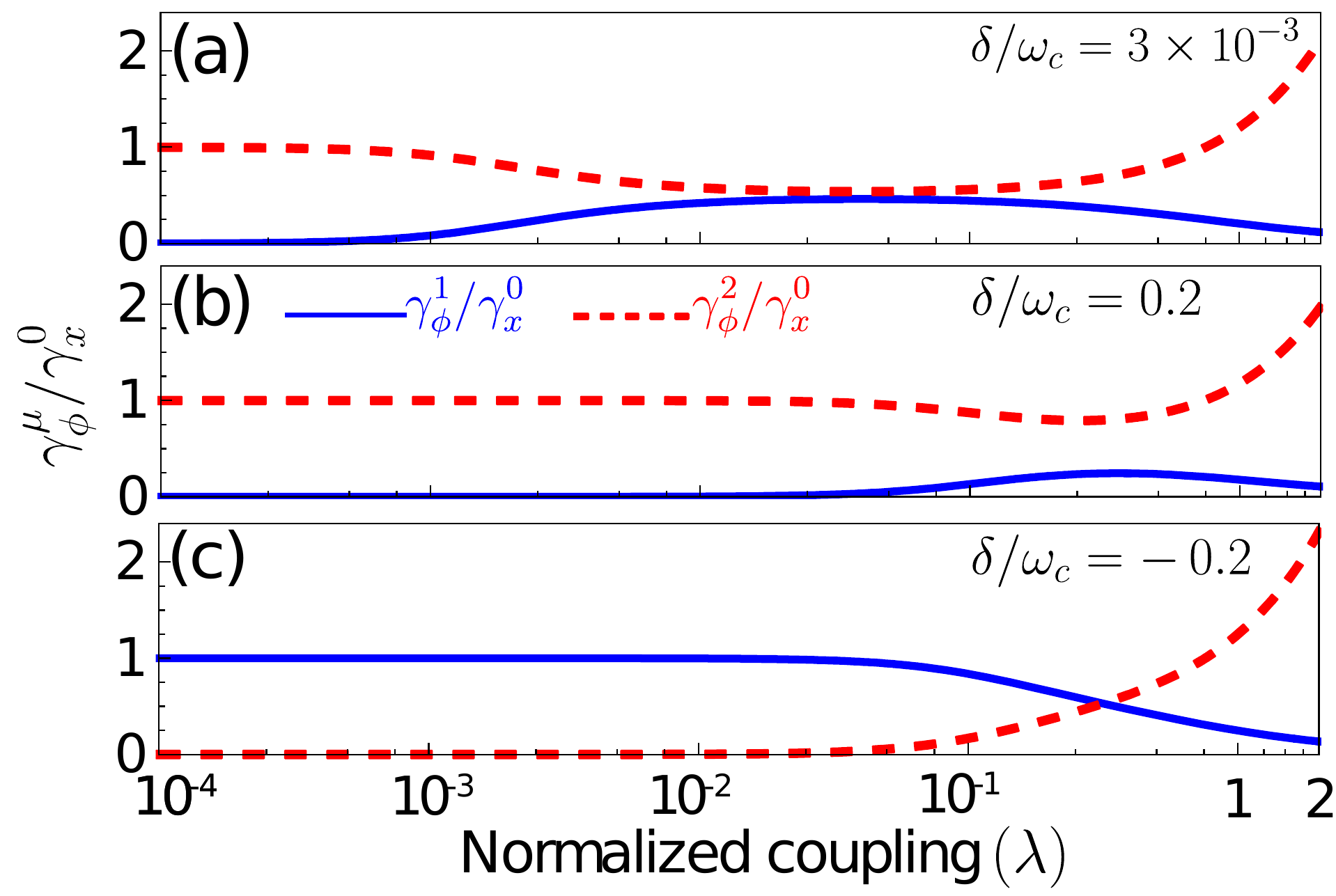}
    \caption {Hopfield model. Normalized pure dephasing rate of the lower and upper polaritons, originating from exciton dephasing, versus the normalized coupling strength, obtained for different exciton-cavity detunings, and considering only the matter pure dephasing. }
    \label{fig:dip}
\end{figure}

\figurepanel{fig:Cav_stra}{a} displays the normalized pure dephasing rate $\gamma_{\tilde 1'_\pm, \tilde 0}/\gamma^0_\phi$ for the two lowest energy transitions, considering a small qubit-cavity detuning $\delta = 3 \times 10^{-3}$ and in the case of only qubit pure dephasing. In the limit of negligible coupling strength, where $| \tilde 1'_+ \rangle \to |e,0 \rangle$ and  $| \tilde 1'_- \rangle \to |g,1 \rangle$, the  standard results are recovered, and only $(\tilde 1'_-, \tilde 0')$ is affected by the qubit pure dephasing. When the coupling becomes comparable to the detuning, as expected, pure dephasing is shared among the two transitions, since the energy eigenstates $| \tilde 1'_\pm \rangle$ tend to become an equally weighted superposition of $|e,0 \rangle$ and $|g,1 \rangle$. For the normalized coupling strengths $\eta > 0.1$ (the USC regime), pure dephasing becomes less effective for the transition $(\tilde 1'_-, \tilde 0)$, until at stronger couplings (the DSC regime), both the transitions tend to become dephasing free. This behavior reflects the fact  that, when the coupling rate is larger than the bare qubit frequency, a fluctuation at the qubit resonance frequency can have a very low impact on the dressed-state energies. On the contrary, \figpanel{fig:Cav_stra}{b} shows a wrong large pure dephasing rate for the lowest energy transition.
Analogous calculations can be carried out for the case of cavity pure dephasing \cite{SupMat}.

{\em Hopfield model.---}A similar analysis can be carried out for polaritons. We consider the simplest version of the Hopfield model \cite{Hopfield1958}, describing the interaction of a single-mode electromagnetic resonator with a bosonic matter field [with the bosonic annihilation $\hat b$ and creation $\hat b^\dag$ operators] modeling some kind of collective matter excitations. The system Hamiltonian in the dipole gauge reads
\be
\label{eq: H Hopfield dipole}
\hat H_{\rm D} =
 \hat{H}_0 + i  \lambda \,\omega_c (\hat{a}^\dag - \hat{a}) (\hat b + \hat b^\dag ) + \omega_c\, \lambda^2 \,(\hat b + \hat b^\dag)^2\,  ,
\ee
where $\hat{H}_0 = \omega_c \hat{a}^\dag \hat{a} +  \omega_x \hat b^\dag \hat b$, and $\lambda$ is the normalized coupling strength. An equivalent model can be obtained in the Coulomb gauge \cite{Garziano2020}:
\be
\label{eq: H Hopfield Coulomb}
\hat H_{C} =    \hat{H}_0 -  i   \omega_x \lambda\,(\hat{b}^\dag - \hat{b})(\hat{a}^\dag + \hat{a}) +   {\cal D} (\hat{a}^\dag + \hat{a})^2\, ,
\ee
where ${\cal D} = \omega_x \lambda^2$.
The two Hamiltonians are related by the unitary gauge transformation $H_D = \hat{{T}} \hat H_C \hat{{T}}^\dag$, where 
$\hat{{T}}= {\rm exp}[-i \lambda (\hat a + \hat a^\dag)(\hat b + \hat b^\dag)]$.

As well known, the interaction gives rise to polaritonic resonances, which results from the mixing of the two bosonic modes. It is possible to diagonalize the system expressing the photon and exciton operators in terms of  polaritonic (bosonic) operators \cite{Hopfield1958}.
For $\mu=1,2$ (lower and upper polariton, respectively), we have
\be
\hat y= \sum_{\mu =1}^2 \left(U_y^{\mu} P_{\mu} - \ V_y^{\mu} P^\dag_{\mu}\right) \, , \quad (y=a,b) \, .
\ee
The diagonalization procedure determines both polariton eigenfrequencies $\Omega_\mu$, which are gauge invariant, and the Hopfield coefficients, which are gauge dependent. As a consequence, also the polariton operators are gauge dependent. We use primed operators and coefficients for the dipole gauge. 

By neglecting issues related to the light-matter interaction, dephasing effects can be modeled by introducing the perturbation Hamiltonian,
\be\label{pdPW}
\hat V_{\rm dep}(t) = f_c(t) \hat a^\dag \hat a + f_x(t) \hat b^\dag \hat b\, ,
\ee
describing the stochastic fluctuation of the resonance frequencies of the components.
Following the reasoning of the previous section, when including the light-matter interaction, it turns out that \eqref{pdPW} is incorrect, and its corrected form is gauge dependent:
\be\label{pdPCd}
\hat V^D_{\rm dep}(t) = f_c(t) \hat a_D^\dag \hat a_D + f_x(t) \hat b^\dag \hat b\, ,
\ee
\be\label{pdPCC}
\hat V^C_{\rm dep}(t) = f_c(t) \hat a^\dag \hat a + f_x(t) \hat b_C^\dag \hat b_C\, ,
\ee
where $\hat a_D = \hat{{T}} \hat a \hat{{T}}^\dag = \hat a + i \lambda (\hat b + \hat b^\dag)$ and $\hat b_C = \hat{{T}}^\dag \hat b \hat{{T}} = \hat b - i \lambda (\hat a + \hat a^\dag)$. Notice that here $\hat a_D$ ($\hat b_C$) is the {\em physical} photonic (excitonic) annihilation operator in the dipole (Coulomb) gauge. By {\em physical}, we mean the operators that describe the annihilation of the physical quanta of the fields \cite{{Garziano2020}}. 
The polariton pure dephasing rates can be obtained by expanding Eqs. (\ref{pdPCd}) and (\ref{pdPCC}) in terms of the polariton operators, and then applying the standard master equation method to obtain the Lindbladian terms, in analogy with the results of the previous section \cite{SupMat}. From the obtained master equation, the equations of motion for the mean values of the polariton operators are $\partial_t \langle {\hat P}_\mu \rangle = (-i \Omega_\mu- \gamma^\mu_\phi/2) \langle {\hat P}_\mu \rangle$, where
\be\label{gp}
\gamma^\mu_\phi = \gamma^0_c \left( |U^\mu_a|^2 + |V^\mu_a|^2\right)  + \gamma^0_x \left(|U^{\mu\, \prime}_b|^2 + |V^{\mu\, \prime}_b|^2\right)\, .
\ee

\begin{figure}[t]
    \centering
    \includegraphics[width= 1\linewidth]{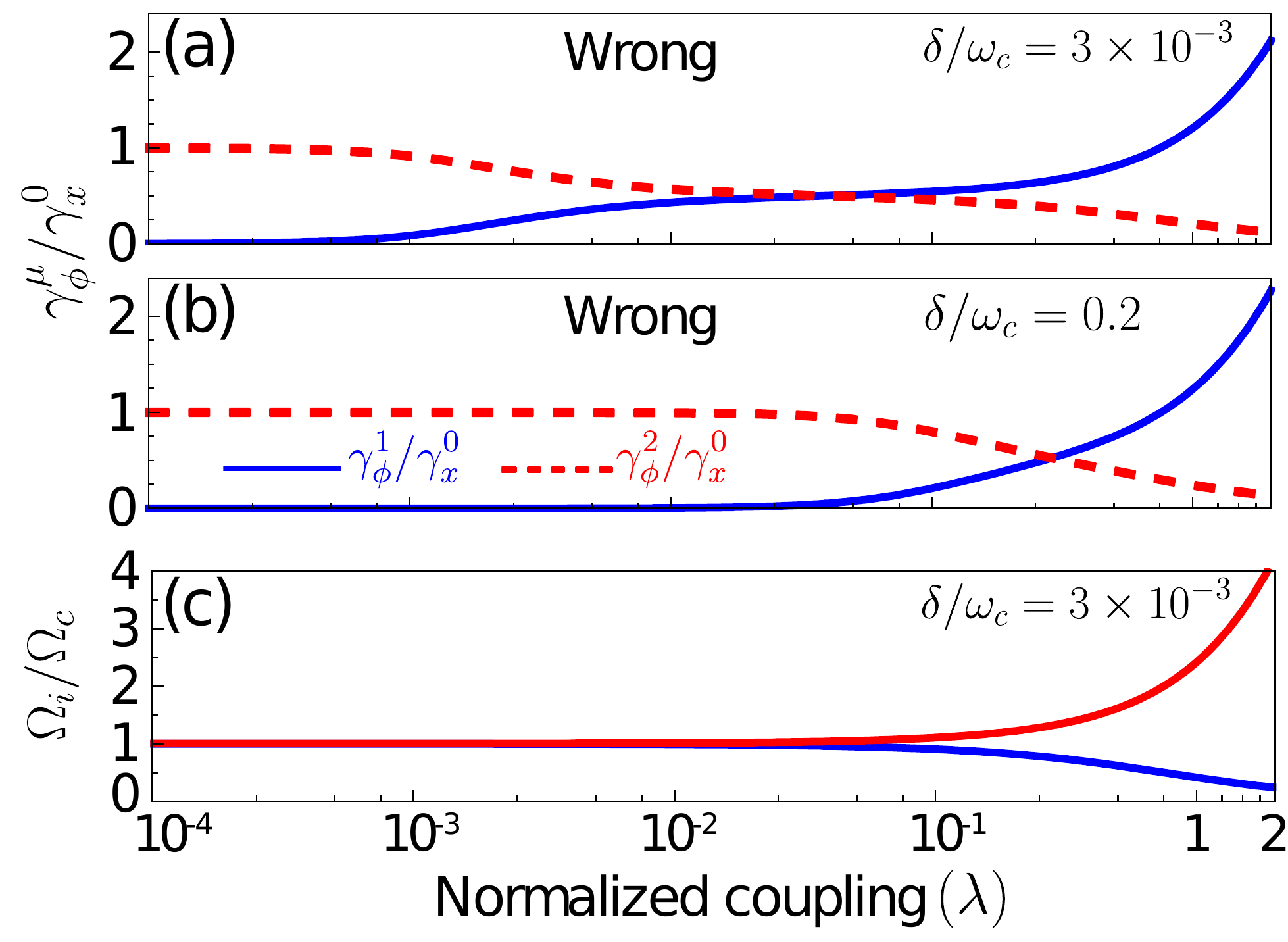}
    \caption {
   Hopfield model. Wrong (see text) normalized pure dephasing rates of the lower and upper polaritons, originating from exciton dephasing, versus the normalized coupling strength, obtained for two different exciton-cavity detunings, and considering only the matter pure dephasing (a, b). Panel (c): frequencies of the two polariton modes for a qubit-cavity detuning $\delta / \omega_c = 3 \times 10^{-3}$.}
    \label{fig:coul}
\end{figure}
This result can be very different from what could be obtained starting from \eqref{pdPW} (the Coulomb gauge), or from $\hat V'_{\rm dep}$ obtained from \eqref{pdPW} replacing all the not-primed operators with primed ones.

\figurepanel{fig:dip}{a} shows the normalized pure dephasing rates for the two polariton modes ($\gamma^\mu_\phi / \gamma^0_x$), for the case of the zero photonic noise ($\gamma_c^0 = 0$),  and considering three different values of the exciton-cavity detuning $\delta$. We observe that, at large coupling rates, independently of the detuning, the lower polariton dephasing rate tends to zero. This effect is a direct consequence of the fact that the lower polariton resonance frequency tends rapidly to zero for $\lambda \to \infty$ [see \figpanel{fig:coul}{c}], independently of the detuning. This implies that any small fluctuation of the resonance frequencies of the components does not induce fluctuations and, hence, dephasing in the polariton mode. For comparison, \figpanels{fig:coul}{a}{b} display the wrong result $\gamma^\mu_\phi / \gamma^0_x=|U^{\mu}_b|^2 + |V^{\mu}_b|^2$, obtained by neglecting the changes of the  form of subsystems-observables, which can be induced by the interaction, as calculated for two different detunings. Evident differences emerge when entering the USC regime with $\lambda \sim 0.1$. Moreover, at larger coupling rates, in the DSC regime, the behavior of the lower and upper polaritons is clearly inverted.

{\em Conclusions.---} We have shown how to calculate correctly the pure dephasing rate in cavity QED systems, considering two prototypical models: the QRM and the Hopfield model. In the latter model, we found that pure dephasing effects in the lower polariton branch tend to be reduced in the USC regime, and tend to get suppressed increasing further the coupling [see \figpanel{fig:dip}{a}]. On the contrary, the influence of pure dephasing increases at increasing coupling strengths for upper polaritons. We hope that these results, closely connected to the gauge principle, will stimulate experimental tests for various polariton systems, where these interaction regimes have been observed \cite{Kockum2018}. In a number of experiments, it was observed that the upper polariton clearly displays a larger line broadening with respect to the lower one \cite{Scalari2012, Gambino2014, Bayer2017, Rajabalipolaritons019} in agreement with the results presented here. However, since in these systems different broadening mechanisms enter into play, further investigations are required.

{\em Acknowledgments.---}
\noindent S.S. acknowledges the Army Research Office (ARO)
(Grant No. W911NF1910065).

\noindent A.M. and S.A. are supported by the Polish National Science
Centre (NCN) under the Maestro Grant No. DEC-2019/34/A/ST2/00081.
S.A. was also supported by the grant No. POWER.03.05.00-00-Z303/17
within the University of Tomorrow Project of Adam Mickiewicz University.

\bibliography{pra}

\clearpage
\onecolumngrid
\setcounter{equation}{0}
\setcounter{figure}{0}
\setcounter{table}{0}
\setcounter{page}{1}
\makeatletter
\renewcommand{\theequation}{S\arabic{equation}}
\renewcommand{\thefigure}{S\arabic{figure}}
\renewcommand{\bibnumfmt}[1]{[S#1]}
\renewcommand{\citenumfont}[1]{S#1}
\renewcommand{\thepage}{S\arabic{page}}

\begin{center}
\textbf{\huge Supplemental Material}
\end{center}

\begin{table}[h]
 \begin{tabular}{ccc}
 \hline\hline
 Hamiltonians \& operators & Coulomb gauge         & Dipole gauge
\\
\hline
 Rabi  Hamiltonian
 & $\hat {\cal H}_C$ in \eqref{eq:H Coulomb}
 & $\hat {\cal H}_D = {\cal T} \hat H_C {\cal T}^\dagger$ above \eqref{eq:H Coulomb}
\\
&
& with ${\cal T}=\exp[-i\hat{\cal A}\hat \sigma_x]$
\\
Rabi annihilation operators
& $\hat a_C = \hat a$
& $\hat a_D = {\cal T} \hat a {\cal T}^\dagger=\hat a+i\eta\hat \sigma_x$ below \eqref{eq: V_dep in dipole}
\\
Rabi perturbation Hamiltonian &&
\\
for pure dephasing
& $\hat {\cal V}_{\phi}^C$ in \eqref{eq: V_dep in Coulomb}
& $\hat {\cal V}_{\phi}^D$ below \eqref{eq:H Coulomb}
\\
\hline
Hopfield  Hamiltonian
& $\hat H_C$ in \eqref{eq: H Hopfield Coulomb}
& $\hat H_D = {\cal T} \hat H_C {\cal T}^\dagger$ in \eqref{eq: H Hopfield dipole}
\\
&&
with ${\cal T}=\exp[-i\lambda(\hat a+\hat a^\dagger)(\hat b+\hat b^\dagger)]$
\\
Hopfield annihilation operators
& $\hat a_C = \hat a$
& $\hat a_D={\cal T} \hat a {\cal T}^\dagger=\hat a+i\lambda(\hat b+\hat b^\dagger)$ below \eqref{pdPCC}
\\
& $\hat b_C={\cal T}^\dagger \hat a {\cal T}=\hat b-i\lambda(\hat a+\hat a^\dagger)$ below \eqref{pdPCC}
& $\hat b_D= \hat{b}$
\\
Hopfield perturbation Hamiltonian &&
\\
for pure dephasing
& $\hat V_{\rm dep}^C$ in \eqref{pdPCC}
& $\hat V_{\rm dep}^D$ in \eqref{pdPCd}
\\
 \hline\hline
\end{tabular}
\caption{Comparison of the basic formulas in the Coulomb and dipole gauges
for the quantum Rabi and Hopfield models.}
 \label{table1}
\end{table}

\section{Pure dephasing in the quantum Rabi model}
\label{app: Rabi model}

Here we analyze how to describe the correct and gauge invariant pure dephasing effects in the quantum Rabi model (QRM), following the procedure described in Ref.~\cite{Beaudoin2011} and considering both cavity and qubit decoherence. We start by considering the quantum Rabi Hamiltonian with an additional zero-mean stochastic modulation of the qubit resonance frequency $\hat{\mathcal{V}}_{\rm dep}^{q} = f_q (t) \hat{\sigma}_z$. Expressing the Hamiltonian in the dressed basis and moving to the interaction picture with respect to $\hat{\mathcal{V}}_{\rm dep}^{q}$, we obtain

\be
\hat{\mathcal{V}}_{\rm dep}^{q} (t) = f(t) \sum_{j,k} \mel{j}{\hat{\sigma}_z}{k} \dyad{j}{k} e^{i \omega_{jk} t}~,
\ee
where $\ket{j}$ are the eigenstates of the total Hamiltonian and $\omega_{jk}$ are the transition frequencies. Expressing $f(t)$ in terms of its Fourier decomposition, and assuming that the main contribution to dephasing results from a small frequency interval around $\omega_{jk}$ \cite{Beaudoin2011}, we obtain

\be
\label{eq_app: qubit dephasing term interaction picture}
\hat{\mathcal{V}}_{\rm dep}^{q} (t) = \sum_{j,k} \hat{\sigma}_z^{jk} \dyad{j}{k} f_{-\omega_{jk}}(t)~,
\ee
where
\be
\label{eq_app: f_omega (t)}
f_{\omega_{jk}}(t)=\sqrt{S_f(\omega_{jk})}\xi_{\omega_{jk}}(t)~,
\ee
$S_f(\omega)$ is the spectral density of $f(t)$, and $\xi(\omega)$ such that $\langle \xi(\omega) \rangle = 0$ and $\langle \xi (\omega) \xi( \omega ') \rangle = \delta (\omega - \omega ')$ (i.e., corresponding to white noise). If the transition frequencies $\omega_{jk}$ are well-separated, we can treat each term of the above summation as an independent noise \cite{Beaudoin2011}. 

We are now able to write down the dressed Lindbladian in case of qubit pure dephasing:
\be
 \label{lindbladiandr}
	\mathcal L_{\rm dr}
	\boldsymbol{\cdot}=\mathcal{D}\left[\sum_j\Phi^{j}\dyad{j}{j}\right]\boldsymbol{\cdot}+\sum_{j, k\neq j}\Gamma_\phi^{jk}\,\mathcal{D}\left[\ket{j}\bra{k}\right]\boldsymbol{\cdot}~,
\ee
where 
\be	\label{eqn:phi_j}
	\Phi^j=\sqrt{\frac{\gamma_\phi(0)}{2}}\sz^{jj},
\ee
and
\be\label{eq:GammaDressedDephasing}
	\Gamma_\phi^{jk}=\frac{\gamma_\phi(\omega_{kj})}{2} \left|\sz^{jk}\right|^2~.
\ee
The whole procedure described above can also be applied to the case of cavity pure dephasing, by considering the QRM Hamiltonian with an additional zero-mean stochastic modulation of the cavity resonance frequency $\hat{\mathcal{V}}_{\rm dep}^{c} = f_c (t) \adop \aop$. In this case, this stochastic perturbation, expressed in the dressed basis and in the interaction picture, becomes
\be
\hat{\mathcal{V}}_{\rm dep}^{c} (t) = \sum_{j,k} \mel{j}{\adop \aop}{k} \dyad{j}{k} f_{-\omega_{jk}}(t)~,
\ee
while the Lindbladian remains in the same form of \eqref{lindbladiandr}, with the only difference of $\Phi^j$ and $\Gamma_\phi^{jk}$, which become respectively,
\bea
\Phi^j &=& \sqrt{\frac{\gamma_\phi(0)}{2}} \mel{j}{\adop \aop}{j} \, , \\
\Gamma_\phi^{jk} &=& \frac{\gamma_\phi(\omega_{kj})}{2} \left| \mel{j}{\adop \aop}{k} \right|^2~.
\eea
However, we have seen in the main text that the approach described above does not reproduce the correct results. In particular, we have shown that, if one uses the Coulomb or dipole gauge, significantly different results can be obtained. For example, when using the Coulomb gauge, the bare $\hat{\sigma}_z$ operator becomes $\hat{\sigma}_z^{C} = \hat{\mathcal{T}}^\dag \hat{\sigma}_z \hat{\mathcal{T}}$, since the minimal coupling is applied to the matter system, while the photonic operator $\adop \aop$ becomes $\adop_D \aop_D = \hat{\mathcal{T}} \adop \aop \hat{\mathcal{T}}^\dag$ in the dipole gauge. Thus, to correctly describe pure dephasing effects, we need to substitute in the Lindbladian given in \eqref{lindbladiandr}: $\hat{\sigma}_z \to \hat{\sigma}_z^{C}$ in the Coulomb gauge, and $\adop \aop \to \adop_D \aop_D$ in the dipole gauge.

\subsection{Analytical derivation of the pure dephasing rates}
By adopting the procedure described above, we are able to derive analytically the pure dephasing rates of both cavity and qubit. Starting from the Coulomb gauge and using \eqref{lindbladiandr}, we discard the off-diagonal terms $\Gamma_\phi^{jk}$ since this contribution is significant only if the dephasing bath has a spectral weight at the potentially high frequency $\omega_{jk}$, leading to the following equation:
\be
\dot{\hat{\rho}} = -i \comm{\hat{\mathcal{H}}_C}{\hat{\rho}} + \frac{\gamma_\phi (0)}{2} \mathcal{D} \left[ \sum_j  \sigma_z^{C, jj} \dyad{j}{j} \right] \hat{\rho}\, ,
\ee
where $\sigma_z^{C, jj} = \mel{j}{\hat{\sigma}_z}{j}$. We now expand the Lindblad dissipator
\bea
\mathcal{D} \left[ \sum_j \sigma_z^{C, jj} \dyad{j}{j} \right] \hat{\rho} &=& \frac{1}{2} \left[ 2 \sum_j \sum_{j'} \sigma_z^{C, jj} \sigma_z^{C, j'j'} \dyad{j}{j} \hat{\rho} \dyad{j'}{j'} - \sum_j \sum_{j'} \sigma_z^{C, jj} \sigma_z^{C, j'j'} \ket{j'}\bra{j'}\ket{j}\bra{j} \hat{\rho} \right. \\
&&- \left. \sum_j \sum_{j'} \sigma_z^{C, jj} \sigma_z^{C, j'j'} \hat{\rho} \ket{j'}\bra{j'}\ket{j}\bra{j} \right] \, ,
\eea
and we focus on the matrix element of the density matrix relative to the transition $(\tilde{1}_{-}, \tilde{0})$, but the same procedure can be applied to all the other transitions. The corresponding equation (in the interaction picture) for that matrix element becomes
\bea
\frac{d}{d t} \hat{\rho}^{(I)}_{\tilde{1}_{-},\tilde{0}} &=& \frac{\gamma_\phi (0)}{4} \bra{\tilde 1_{-}}\left[2\sum_j\sum_{j^\prime} \sigma_z^{C, jj}\sigma_z^{C, j'j'} \ket{j}\bra{j}\hat{\rho}^{(I)}\ket{j^\prime}\bra{j^\prime}-\sum_j|\sigma_z^{C, jj}|^2\ket{j}\bra{j}\hat{\rho}^{(I)}-\sum_j|\sigma_z^{C, jj}|^2\hat{\rho}^{(I)} \ket{j}\bra{j}\right]\ket{\tilde 0}\nonumber\\
&=& \frac{\gamma_\phi (0)}{4} \left[2\sum_j\sum_{j^\prime} \sigma_z^{C, jj}\sigma_z^{C, j'j'} \bra{\tilde 1_{-}}\ket{j}\bra{j}\hat{\rho}^{(I)}\ket{j^\prime}\bra{j^\prime}\ket{\tilde 0} - \sum_j|\sigma_z^{C, jj}|^2\bra{\tilde 1_{-}}\ket{j}\bra{j}\hat{\rho}^{(I)}\ket{\tilde 0} \right. \nonumber \\
&&- \left. \sum_j|\sigma_z^{C, jj}|^2 \bra{\tilde 1_{-}}\hat{\rho}^{(I)}\ket{j}\bra{j}\ket{\tilde 0}\right]\nonumber\\
&=& \frac{\gamma_\phi (0)}{4} \left[2 \sigma_z^{C, \tilde{1}_- \tilde{1}_-}\sigma_z^{C, \tilde{0} \tilde{0}} \bra{\tilde 1_{-}}\hat{\rho}^{(I)}\ket{\tilde 0}-|\sigma_z^{C, \tilde{1}_- \tilde{1}_-}|^2\bra{\tilde 1_{-}}\hat{\rho}^{(I)}\ket{\tilde 0}-|\sigma_z^{C, \tilde{0} \tilde{0}}|^2 \bra{\tilde 1_{-}}\hat{\rho}^{(I)}\ket{\tilde 0}\right] \nonumber \\
&=& - \frac{\gamma_\phi (0)}{4} \abs{ \sigma_z^{C, \tilde{1}_- \tilde{1}_-} - \sigma_z^{C, \tilde{0} \tilde{0}} }^2 \hat{\rho}^{(I)}_{\tilde{1}_{-},\tilde{0}} \, .
\eea
By choosing the dipole gauge, one should replace $\sigma_z^{C, jj} \to \sigma_z^{jj}$.
The same procedure is valid also for cavity pre dephasing, where we need to use $\adop \aop$ in the Coulomb gauge and $\adop_D \aop_D$ in the dipole gauge.

\section {Pure dephasing in bosonic systems}
We now consider pure dephasing effects in bosonic systems. 
First, we consider a simple non-interacting harmonic oscillator, then we analyze the Hopfield model.

\subsection{Non-interacting harmonic oscillator}\label{nho}
Here we consider a single-mode bosonic field described by the harmonic oscillator Hamiltonian $\hat{H}_0 = \omega_0 \adop \aop$ affected by pure dephasing. Analogously to what we described in previous sections, in order to consider the dephasing effects, we introduce an additional zero-mean stochastic modulation of the resonance frequency $\hat{\mathcal{V}}_{\rm dep}^{h} = f_h (t) \adop \aop$. Moving to the interaction picture, we notice that this component does not rotate, since it has a zero-frequency oscillation. Thus, transforming $f_h (t)$ in its Fourier components, and assuming that the main contribution to dephasing comes from a small frequency interval around $\omega = 0$ \cite{Beaudoin2011}, we obtain 
\be
\hat{{V}}_{\rm dep}^{h} (t) = f_0 (t) \adop \aop~,
\ee
where $f_0 (t) = \sqrt{S_f(0)} \xi_0(t)$. This equation is quite similar to \eqref{eq_app: qubit dephasing term interaction picture} with the only difference that here we do not have the expansion in the dressed basis (since we are not considering a hybrid quantum system), and that we have only the zero-frequency contribution (since $\hat{\mathcal{V}}_{\rm dep}^{h}$ rotates at zero frequency in the interaction picture). These considerations allow us to write the Lindbladian describing this pure dephasing effect as
\be
 \label{lindbladianaadag}
	\mathcal L
	\boldsymbol{\cdot}=\sqrt{\frac{\gamma_\phi(0)}{2}}\mathcal{D}\left[\hat a^\dag \hat a \right]\boldsymbol{\cdot} \,,
\ee
with $\gamma_\phi (0) = 2 S_f(0)$.

\subsection{Hopfield model}\label{app: hopfield}
Here we analyze pure dephasing effects in the Hopfield model, following the procedure described in the previous sections and extending the results of Ref.~\cite{Beaudoin2011}. Moreover, we  consider both light and matter decoherence. First, it is useful to diagonalize the Hopfield Hamiltonian using the polaritonic operators \cite{Hopfield1958}, where the lower and upper polariton operators ($\mu=1,2$) can be defined as
\be\label{pol}
\hat P^\mu= U^\mu_{ b}\hat b + U^\mu_{ a}\hat a +
V^\mu_{ b}\hat b^\dag + V^\mu_{ a}\hat a^\dag\,.
\ee
Using the property
\be\label{pro}
|U^\mu_{ b}|^2+ |U^\mu_{ a}|^2 -
|V^\mu_{b}|^2 - |V^\mu_{ a}|^2=1\,,
\ee
which guarantee the correct polariton commutation rules \cite{Hopfield1958}, we can invert \eqref{pol} in order to obtain
\begin{subequations}
\begin{align}
\hat a&= \sum_{\mu =1}^2 \left(U^\mu_{ a} P^{}_{\mu} - V^\mu_{ a} P^{\dag}_{\mu}\right)\, , \\
\hat b&= \sum_{\mu =1}^2 \left(U^\mu_{ b} P^{}_{\mu} - V^\mu_{ b} P^\dag_{\mu}\right)\, .
\end{align}
\end{subequations}
To describe the matter pure dephasing, we consider an additional zero-mean stochastic modulation of the matter resonance frequency $\hat{V}_{\rm dep}^{x} = f_x (t) \hat{b}^\dag \hat{b}$. In terms of the polaritonic operators we have
\be
\hat{b}^\dag \hat{b} = A_1 \hat{P}_1^\dag \hat{P}_1 + A_2 \hat{P}_2^\dag \hat{P}_2 + B_{12} \hat{P}_1^\dag \hat{P}_2 + B_{21} \hat{P}_2^\dag \hat{P}_1 \, ,
\ee
with
\bea
A_\mu &=& \abs{U_b^\mu}^2 + \abs{V_b^\mu}^2 \\
B_{12} &=& B_{21}^* = U_b^{1 \, *} U_b^2 + V_b^1 V_b^{2 \, *} \, ,
\eea
where we have included only the terms which do not oscillate in time, or oscillate at low frequency, corresponding to applying the rotating wave approximation (RWA), and we have eliminated the constants derived from commutation rules, which have no dynamical consequences. Moving to the interaction picture, this contribution becomes
\be
\label{eq_app: V dephasing matter in interaction picture}
\hat{V}_{\rm dep}^{x} (t) = f_x(t) \left[ A_1 {\hat P}_1^\dag {\hat P}_1 + A_2 {\hat P}_{2}^\dag {\hat P}_2 \right. + \left. e^{-i \omega_{21} t} B_{12} {\hat P}_1^\dag {\hat P}_2 + e^{i \omega_{21} t} B_{21} {\hat P}_2^\dag {\hat P}_1 \right] \, ,
\ee
where $\omega_{21} = \omega_2 - \omega_1$ with the polaritonic eigenfrequencies $\omega_i$. \eqaref{eq_app: V dephasing matter in interaction picture} can be written in a more compact form as
\[\hat{V}_{\rm dep}^{x} = f_x(t) \left[ \hat{D}_{12} + e^{-i \omega_{21} t} \hat{M}_{12} + e^{i \omega_{21} t} \hat{M}_{12}^\dag \right] \, ,
\]
with
\bea
\hat{D}_{12} &=& A_1 \hat{P}_1^\dag \hat{P}_1 + A_2 \hat{P}_2^\dag \hat{P}_2 \, , \\
\hat{M}_{12} &=& B_{12} \hat{P}_1^\dag \hat{P}_2 \, ,
\eea
and using the results presented in the previous sections, we obtain
\be
\hat{V}_{\rm dep}^{x} (t) = f_0 (t) \hat{D}_{12} + f_{\omega_{21}}(t) \hat{M}_{12} + f_{- \omega_{21}} (t) \hat{M}_{12}^\dag \, ,
\ee
with $f_{\omega} (t)$ expressed in \eqref{eq_app: f_omega (t)}. Thus, the resulting Lindbladian in the case of matter pure dephasing is
\be	\label{eq_app: lindbladian matter dephasing without gauge}
	\mathcal{L} \boldsymbol{\cdot} = \frac{1}{2}\gamma_\phi(\omega_{21}) \mathcal{D}[\hat M_{12}]  \boldsymbol{\cdot}+\frac{1}{2}\gamma_\phi(-\omega_{21}) \mathcal{D}[\hat M_{12}^\dag] \boldsymbol{\cdot} + \frac{1}{2}\gamma_\phi(0) \mathcal{D}[\hat D_{12}] \boldsymbol{\cdot}\, ,
\ee
with $\gamma_\phi (\omega) = 2 S_f(\omega)$.

The same procedure, as described above, can also be applied to the case of cavity pure dephasing, by considering an additional zero-mean stochastic modulation of the cavity resonance frequency $\hat{V}_{\rm dep}^{c} = f_c (t) \adop \aop$. The procedure remains the same for the matter dephasing case, except that now we consider
\be
\hat{a}^\dag \hat{a} = A_1 \hat{P}_1^\dag \hat{P}_1 + A_2 \hat{P}_2^\dag \hat{P}_2 + B_{12} \hat{P}_1^\dag \hat{P}_2 + B_{21} \hat{P}_2^\dag \hat{P}_1 \, ,
\ee
where
\bea
\label{eq_app: A_mu cavity}
A_\mu &=& \abs{U_a^\mu}^2 + \abs{V_a^\mu}^2 \, , \\
\label{eq_app: B_12 cavity}
B_{12} &=& B_{21}^* = U_a^{1 \, *} U_a^2 + V_a^1 V_a^{2 \, *} \, .
\eea
This yields a Lindbldian of the same form of \eqref{eq_app: lindbladian matter dephasing without gauge} with the only difference for the polariton coefficients expressed in Eqs.~(\ref{eq_app: A_mu cavity}) and (\ref{eq_app: B_12 cavity}).

However, we have seen in the main text that this approach can lead to wrong results, depending on the chosen gauge. Indeed, when using the Coulomb gauge, the matter operator $\hat{b}$ becomes $\hat{b}_C = \hat{{T}}^\dag \hat{b} \hat{{T}}$, since the minimal coupling is applied to the matter system. On the contrary, when using the dipole gauge, the minimal coupling is applied to the photonic system, and the \emph{dressed} photonic operator becomes $\aop_D = \hat{{T}} \aop \hat{{T}}^\dag$. This consideration leads us to note that the polariton diagonalization leads to different Hopfield coefficients if we choose the Coulomb or dipole gauge. In particular, in the dipole gauge, we have 
\be
\hat b=\sum_{\mu =1}^2 \left(U_b^{\mu \prime} P^\prime_{\mu} - \ V_b^{\mu \prime} P^{\prime \dag}_{\mu}\right) \, ,
\ee
where $P^\prime_{\mu}$ are the polariton operators obtained by diagonalizing the Hopfield Hamiltonian in the dipole gauge. While in the Coulomb gauge we have
\bea\label{eq:bC}
\hat b_C&=& \hat{ T}^\dag \left[ \sum_{\mu =1}^2 \left(U_b^{\mu \prime} P^\prime_{\mu} - \ V_b^{\mu \prime} P^{\prime \dag}_{\mu}\right)  \right]\hat{ T}\nonumber\\
&=&\sum_{\mu =1}^2 \left(U_b^{\mu \prime} \hat{ T}^\dag P^\prime_{\mu}\hat{ T} - \ V_b^{\mu \prime} \hat{ T}^\dag P^{\prime \dag}_{\mu}\hat{ T}\right)\nonumber \\
&=&\sum_{\mu =1}^2 \left(U_b^{\mu \prime} P^{}_{\mu} - \ V_b^{\mu \prime} P^{\dag}_{\mu}\right)\, ,
\eea
which contains the polariton operators obtained by diagonalizing the Hamiltonian in the Coulomb gauge, but with the same coefficients of the dipole gauge. To obtain \eqref{eq:bC}, we have used the relation
\be
\label{Pgauge}
P^{}_{\mu}=\hat{ T}^\dag P^\prime_{\mu}\hat{ T}\, ,
\ee
which, although intuitively obvious, can be rigorously demonstrated using the definition of  polaritonic operators; in particular, those operators that, each in its specific gauge, enable the diagonalization of the gauge-correspondent Hamiltonian.
For example, we have:
\begin{subequations}
\begin{align}
[P^{}_{\mu}, \hat {\cal H}_C]=\Omega_\mu P^{}_{\mu}\, ,\label{commHc}\\
[P^{\prime}_{\mu}, \hat {\cal H}_D]= \Omega_\mu P^{\prime}_{\mu}\, .\label{commHd}
\end{align}
\end{subequations}
In order to demonstrate \eqref{Pgauge}, we can calculate how \eqref{commHc} transforms from the Coulomb to dipole gauge. Gauge invariance implies that the final result has to be equal to \eqref{commHd}.
We obtain:
\begin{subequations}
\bea\label{gaugetransf}
\hat{T} [P^{}_{\mu}, \hat {H}_C]\hat{T}^\dag
&=&\Omega_\mu \hat{T} P^{}_{\mu}\hat{T}^\dag\,, \\
\label{gaugetransf2}
\hat{T} [P^{}_{\mu}, \hat {H}_C]\hat{T}^\dag&=&\hat{T} (P^{}_{\mu} \hat {H}_C-\hat {H}_C P^{}_{\mu})  \hat{T}^\dag\\
&=&\hat{T} P^{}_{\mu} \hat {H}_C \hat{T}^\dag- \hat{T}\hat {H}_C P^{}_{\mu} \hat{T}^\dag
\nonumber\\
&=& \hat{T} P^{}_{\mu} \hat{T}^\dag\hat{T}\hat {H}_C \hat{T}^\dag- \hat{T}\hat {H}_C  \hat{T}^\dag\hat{T}P^{}_{\mu} \hat{T}^\dag \nonumber\\
&=&\hat{T} P^{}_{\mu} \hat{T}^\dag\hat {H}_D - \hat {H}_D  \hat{T}P^{}_{\mu} \hat{T}^\dag
=[\hat{T} P^{}_{\mu}\hat{T}^\dag,  \hat {H}_D]
\nonumber\,.
\eea
\end{subequations}
Combining the results of Eqs.~(\ref{gaugetransf}) and (\ref{gaugetransf2}), we obtain:
\bea\label{gaugetransf3}
[\hat{T} P^{}_{\mu}\hat{T}^\dag,  \hat {H}_D]
=\Omega_\mu \hat{T} P^{}_{\mu}\hat{T}^\dag\, ,
\eea
which is the definition of the polariton operators $P^{\prime}_{\mu}$ in the dipole-gauge (which are the operators that allow the diagonalization of $\hat {\cal H}_D$) given by \eqref{commHd}. Hence, \eqref{Pgauge} is the correct gauge transformation for the polaritonic operators.

The whole analysis described above can be summarized as follows: in the case of matter pure dephasing, the stochastic perturbation is: $\hat{V}_{\rm dep}^x = f_x(t) \hat{b}^\dag \hat{b}$ in the dipole gauge, and $\hat{V}_{\rm dep}^x = f_x(t) \hat{b}_C^\dag \hat{b}_C$ in the Coulomb gauge, where
\be
\hat b ^\dag \hat b = A_1^\prime {\hat P}_1^{\prime \dag} {\hat P}^\prime
_1 + A_2^\prime {\hat P}_{2}^{\prime \dag} {\hat P}^\prime_2 + B_{12}^\prime {\hat P}_1^{\prime \dag} {\hat P}^\prime_2 +  B_{21}^\prime {\hat P}_2^{\prime \dag}{\hat P}^{\prime}_1
\ee
and
\be
\hat{b}_C^\dag \hat{b}_C = A_1^\prime {\hat P}_1^\dag {\hat P}_1 + A_2^\prime {\hat P}_{2}^\dag {\hat P}_2 + B_{12}^\prime {\hat P}_1^\dag {\hat P}_2 +  B_{21}^\prime {\hat P}_2^\dag {\hat P}_1 \, ,
\ee
with
\bea
\label{eq_app: matter polariton coefficients dipole 1}
A_\mu^\prime &=& \abs{U_b^{\mu \prime}}^2 + \abs{V_b^{\mu \prime}}^2 \, , \\
\label{eq_app: matter polariton coefficients dipole 2}
B_{12}^\prime &=& B_{21}^{\prime \, *} = U_b^{1 \prime \, *} U_b^{2 \prime} + V_b^{1 \prime} V_b^{2 \prime \, *} \, .
\eea
As a result, to correctly describe the matter pure dephasing, we need to use the dipole coefficients, given in Eqs.~(\ref{eq_app: matter polariton coefficients dipole 1}) and (\ref{eq_app: matter polariton coefficients dipole 2}), in the Lindbladian expressed in \eqref{eq_app: lindbladian matter dephasing without gauge}, even when using the Coulomb gauge. On the contrary, for the photonic pure dephasing, the stochastic perturbation is: $\hat{V}_{\rm dep}^c = f_c(t) \adop \aop$ in the Coulomb gauge, and $\hat{V}_{\rm dep}^c = f_c(t) \adop_D \aop_D$ in the dipole gauge. Thus, we need to use the Coulomb polariton coefficients in the Lindbladian even when using the dipole gauge.

\section{Additional Figures}
\begin{figure}[ht]
    \centering
    \includegraphics[width= 0.5\linewidth]{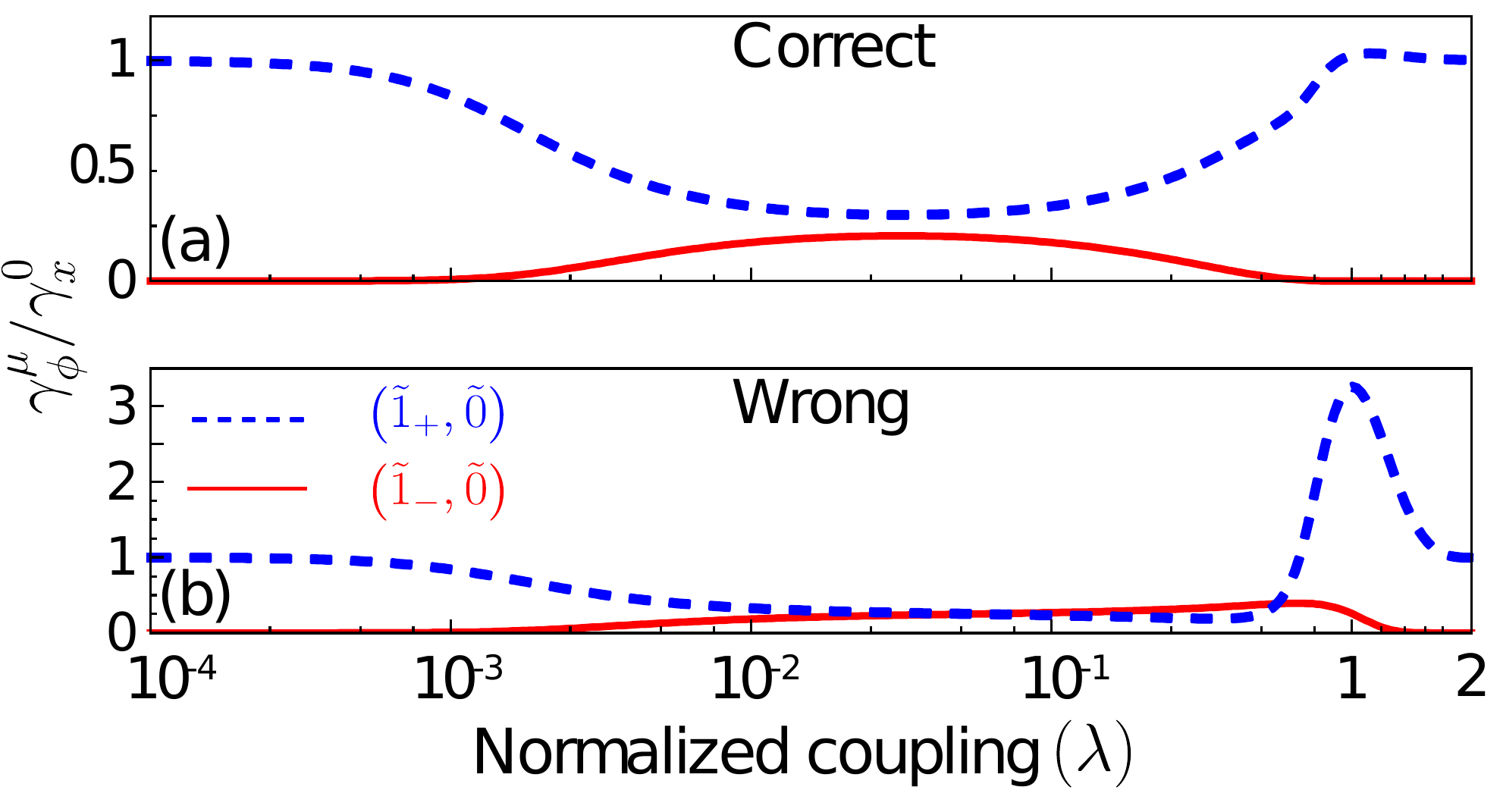}
    \caption {
    Quantum Rabi model: Normalized pure dephasing rate 
    for the two lowest energy transitions, for a small qubit-cavity detuning $\delta = 3 \times 10^{-3}$ assuming only the cavity pure dephasing. (a) Correct gauge-invariant results versus (b) wrong Coulomb gauge results.}
    \label{fig:Cav_str}
\end{figure}
\begin{figure*}[b]
    \centering
    \includegraphics[width= \linewidth]{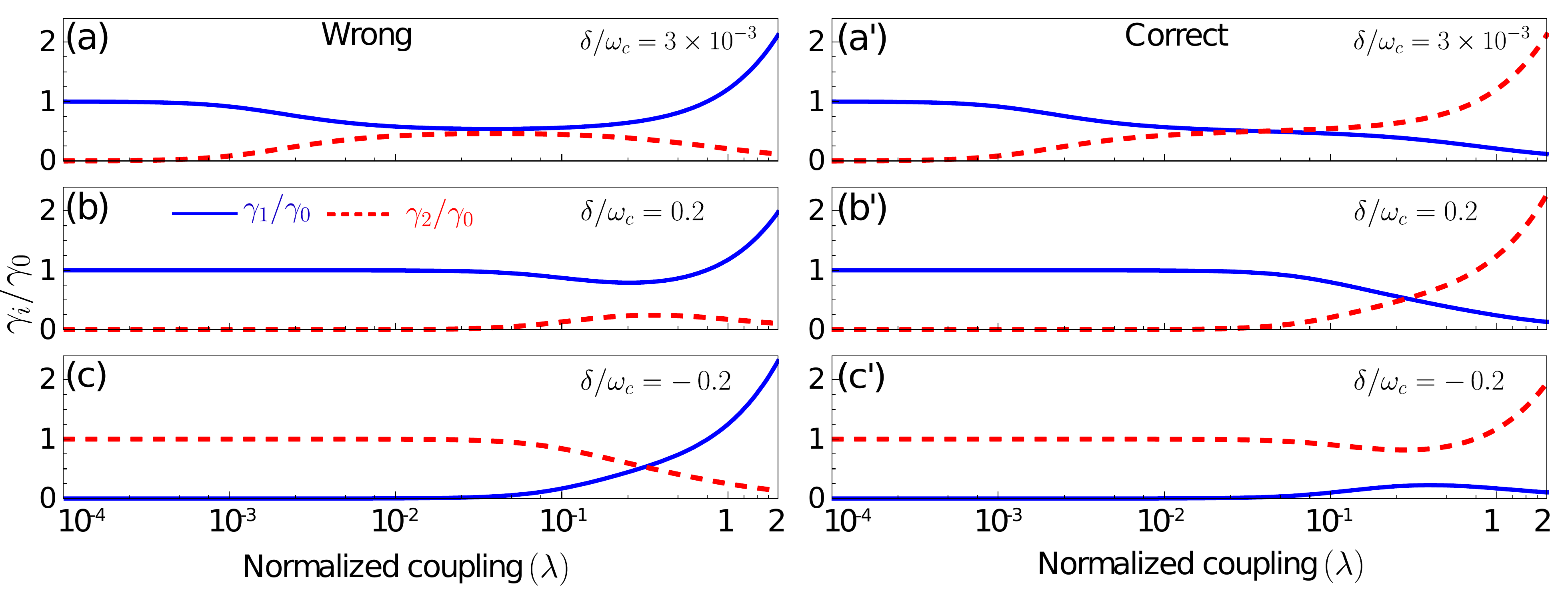}
    \caption {Hopfield model: Pure dephasing rate of the lower and upper polaritons, originating from exciton dephasing, versus the normalized coupling strength, obtained for different exciton-cavity detunings, and considering only cavity pure dephasing.
     }
    \label{fig:Cav_str1}
\end{figure*}

\end{document}